\documentclass[conference]{IEEEtran}
\usepackage{amsmath,cite,bm,amsfonts,amssymb,graphicx,color,mathtools,setspace,comment,multirow,xfrac}
\usepackage{epstopdf}

\DeclareMathOperator{\Var}{Var}
\newcommand{\bH}{\mathbf{H}}
\newcommand{\bh}{\mathbf{h}}
\newcommand{\bG}{\mathbf{G}}
\newcommand{\bg}{\mathbf{g}}
\newcommand{\bI}{\mathbf{I}}
\newcommand{\bW}{\mathbf{W}}
\newcommand{\bD}{\mathbf{D}}
\newcommand{\bE}{\mathbf{E}}
\newcommand{\bR}{\mathbf{R}}
\newcommand{\bxf}{\mathbf{x}_{f}}
\newcommand{\bsf}{\mathbf{s}_{f}}
\newcommand{\bqf}{\mathbf{q}_{f}}
\newcommand{\bwf}{\mathbf{w}_{f}}

\newcommand{\raisecaption}{\vspace{-0.7cm}}

\begin{document}

\title{On the Convergence of Massive MIMO Systems}
\author{\IEEEauthorblockA{Peter J. Smith\IEEEauthorrefmark{1},
											Callum Neil\IEEEauthorrefmark{3},
											Mansoor Shafi\IEEEauthorrefmark{2},
											Pawel A. Dmochowski\IEEEauthorrefmark{3}
											}
\IEEEauthorblockA{\IEEEauthorrefmark{1}
\small{Department of Electrical and Computer Engineering, University of Canterbury, Christchurch, New Zealand}}
\IEEEauthorblockA{\IEEEauthorrefmark{2}
\small{Telecom New Zealand, Wellington, New Zealand}}
\IEEEauthorblockA{\IEEEauthorrefmark{3}
\small{School of Engineering and Computer Science, Victoria University of Wellington, Wellington, New Zealand}}
\IEEEauthorblockA{\small{email:\{pawel.dmochowski,callum.neil\}@ecs.vuw.ac.nz,~p.smith@elec.canterbury.ac.nz,~mansoor.shafi@telecom.co.nz}}}

\maketitle

\begin{abstract}
In this paper we examine convergence properties of massive MIMO systems with the aim of determining the number of antennas required for massive MIMO gains. We consider three characteristics of a channel matrix and study their asymptotic behaviour. Furthermore, we derive ZF SNR and MF SINR for a scenario of unequal receive powers. In our results we include the effects of spatial correlation. We show that the rate of convergence of channel metrics is much slower than that of the ZF/MF precoder properties.
\end{abstract}
\section{Introduction}

Multiple-Input Multiple-Output (MIMO) has been incorporated into emerging wireless broadband standards such as LTE. In general, in time-division duplexing (TDD) mode, as we add more and more antennas to the base station, the performance in terms of data rate, enhanced reliability, improved energy efficiency and interference increases \cite{LARSSON}. Consequently, massive MIMO is an emerging technology, where the number of antennas is scaled up by 1-2 orders of magnitude \cite{MARZETTA,NGO,PITAROKOILIS,HOYDIS,MULLER,MOUSTAKAS,CHUAH} with the aim of even greater performance benefits \cite{LARSSON}.
\par
Motivating the surge in research activities into massive MIMO are the additional gains resulting from a large channel matrix, due to the asymptotics of random matrix theory. For example, the effect of small-scale fading can be averaged out \cite{NGO}, beacuse random elements tend to become deterministic and matrix operations can be computed much easier due to matrices becoming well conditioned \cite{RUSEK}. Consequently as we let the number of antennas tend to infinity, system analysis becomes simplified. 
\par
However, as a very large number of antennas may be impractical due to physical constraints, we would like to determine the point where MIMO systems begin to exhibit these additonal benefits. This is the focus of the paper. The following are the contributions:
\begin{itemize}
	\item We study the convergence properties of the channel matrix by considering three metrics defined in Sec. III.
	\item We investigate precoder properties by deriving new analytical expressions for SNR and SINR of ZF and MF precoders for unequal power gains.
	\item We demonstrate the difference in convergence rate of channel properties and that of precoder characteristics \footnote{Note: Imperfect CSI is considered in future work}.
\end{itemize}


\section{System Model}
\label{sec:SystemModel}


\subsection{System Description}
\label{SystemDescription}
We consider a massive MIMO system with $M$ co-located antennas at the base station serving $K$ single-antenna users. On the down-link (where TDD is assumed) the $K$ terminals collectively receive a $K\times 1$ vector
\begin{equation}
	\bxf = \sqrt{\rho _{f}}\bG^{T}\bsf + \bwf,
\end{equation}
where $\rho _{f}$ is the transmit SNR, $[\cdot]^T$ represents matrix transpose, $\bsf$ is a $M \times 1$ precoded vector of data symbols, $\bqf$, and $\bwf$ is a $K\times 1$ noise vector with independent and identically distributed (iid) $\mathcal{CN}(0,1)$ entries. The transmit power is normalized such that $\mathbb{E}\{ \| \bsf^{2}\| \} = 1$. The $M\times K$ channel matrix, $\bG$, is
\begin{equation}
\label{GHDdef}
	\bG = \bH\bD_{\beta }^{\frac{1}{2}},
\end{equation}
where $\bH$ is a $M\times K$ matrix which accounts for small-scale Rayleigh fading and transmit spatial correlation, and $\bD_{\beta }$ is a diagonal matrix modelling large-scale effects. The diagonal elements of $\bD_{\beta }$ are represented by a $K\times 1$ vector, $\bm{\beta}$, with entries $\beta_j$ representing the link gains.

\subsection{Power Model}
\label{PowerModel}

We consider cases with equal and unequal link gains. The equal power case models a single-user MIMO system where one user has $K$ co-located antennas. This is used as a reference case. While not considered by other authors, the unequal power case models $K$ distributed UEs where each user has a different link gain due to path-loss, shadowing, etc. Since we are considering convergence issues as the system dimension grows large, it is not convenient to generate the link gains using classic log-normal shadowing, path loss models. This is because the variation in the link gains will confound the limiting effects. Also, with random link gains, as the system size increases we may obtain an artificially large number of high gains or small gains. We propose a power model to counter these two problems.
\par
We select the $\beta _{j}$ values from the limiting function $\beta (x)=A\eta ^{x}$, where $\eta $ is arbitrarily set such that $0<\eta <1$, $A=\beta _{\mathrm{max}}$, and $0<x<x_{0}$, such that $x_{0}=\frac{\mathrm{log}(\beta _{\mathrm{min}}/\beta _{\mathrm{max}})}{\mathrm{log}(\eta )}$. This allows us to control the range of the link gains in the interval $[\beta _{\textrm{min}},\beta _{\textrm{max}}]=[A\eta ^{x_{0}},A]$  and also to control the rate of decay of the link gains by $\eta $. Given the parameters $\beta _{\textrm{min}},\beta _{\textrm{max}},\eta $ and the number of users, $K$, the $\beta _{j}$ values are given by $\beta _{j}=\beta (\frac{x_{0}}{2K}(2j-1))$ which gives the $K$ values of $\beta _{1},\beta _{2},\ldots ,\beta _{K}$ as the values of $\beta (x)$ using $K$ values of $x$ evenly spread over $[0,x_{0}]$.

\subsection{Convergence Model}
\label{ConvergenceModel}
Many features of massive MIMO systems are driven by the convergence of various functions as the system size increases. A key issue is the convergence of $\frac{1}{M}\bG^{T}\bG^{\ast }$ to $\bI$. This is examined in Sec. \ref{GGconv} and \ref{ConvergenceMetrics}. Also of interest is the convergence of ZF and MF precoder properties to their asymptotic limits. Thus, in Section \ref{ZFMFconv}, we examine the convergence of the SNR and SINR of ZF and MF precoders, respectively. 

In doing so we will consider the following two types of convergence scenarios:
\begin{enumerate}
	\item $K$ is fixed, $M\rightarrow \infty $,
	\item $\frac{K}{M}=\alpha ^{-1},K\rightarrow \infty $, $\alpha $ is fixed.
\end{enumerate}

\subsection{Correlation Model}

In massive MIMO, if we deploy more and more antennas in a fixed volume, correlation amongst antenna elements will increase. Hence, as we allow the array sizes to grow, we also allow correlations to increase. We consider the Kronecker correlation model, where
\vspace{-5pt}
\begin{equation}
	\bH = \bR_{\mathrm{t}}^{1/2}\bH_{\mathrm{iid}},	\label{Kronecker}
\end{equation}
where $\bH_{\mathrm{iid}}$ is the $M\times K$ iid channel matrix, $\bH $ is the $M\times K$ correlated channel matrix and $\bR_{\mathrm{t}}$ is the $M\times M$ transmit spatial correlation matrix. The elements of $\bR_{\mathrm{t}}$, $r_{ij}$, are based on the exponential correlation matrix model \cite{LOYKA}
\begin{equation}
	r_{ij}=	\rho ^{d_{ij}}, \textrm{ }|\rho |\leq 1, \label{expCorrModel}
\end{equation}
where $d_{ij}$ is the distance between antennas $i$ and $j$, $\rho $ is the correlation decay constant and $r_{ij}$ is the correlation coefficient between antennas $i$ and $j$. We will assume the $M$ colocated antennas are in a Uniform Linear Array (ULA) at the base station.

\vspace{-11pt}
\section{Methodology}
\label{sec:Methodology}

In this section we focus on the second convergence scenario, where both $K$ and $M$ grow large with $\alpha =M/K$ remaining fixed.\footnote{Due to limitations of space all the analysis shown in Sec. \ref{sec:Methodology} is for the iid case, but can be easily modified to include the effects of correlation.}

\vspace{-5pt}
\subsection{Convergence of $\bG^{T}\bG^{\ast}/M$}
\label{GGconv}

First, consider the entries of $\bH^{T}\bH^{\ast }$. Let $\bH=[\bh_{1} \ \bh_{2} \ldots \bh_{K}]$, where $\bh_i$ is an $M \times 1$ column vector with entries, $h_{ij}$, then it is well known that 
\begin{align}
\label{HHMii}
	\left(\frac{\bH ^{T}\bH ^{\ast }}{M}\right)_{ii} = \frac{\bh _{i}^{T}\bh _{i}^{\ast }}{M} = \frac{\sum_{r=1}^{M}{|h_{ir}|^{2}}}{M} = \frac{X}{M},
\end{align}
where $X$ is $\mathcal{X}_{2M}^{2}$ with $2M$ complex degrees of freedom. Similarly
\begin{align}
\label{HHMij}
\left(\frac{\bH ^{T}\bH ^{\ast }}{M}\right)_{ij} = \frac{\bh _{i}^{T}\bh _{j}^{\ast }}{M} = \frac{\sum_{r=1}^{M}{h_{ir}h_{jr}^{\ast }}}{M} = \frac{Y}{M}.
\end{align}
Using the known convergence of $\frac{1}{M}\bH^{T}\bH^{\ast }$ to $\bI$ we have \cite{PAPOULIS}
\begin{align}
	\frac{\bG^{T}\bG^{\ast}}{M} &= \frac{\bD _{\beta }^{\frac{1}{2}}\bH ^{T}\bH ^{\ast }\bD _{\beta }^{\frac{1}{2}}}{M} \rightarrow \bD _{\beta }^{\infty },
\end{align}
where $\bD_{\beta }^{\infty }$ is the limiting value of $\bD_{\beta }$ assuming it exists. The diagonal elements of $\frac{1}{M}\bG^{T}\bG^{\ast }$ are given by
\begin{align}
\label{GGMii}
	\frac{(\bG^{T}\bG^{\ast})_{ii}}{M} = 
	\frac{\beta _{i}X}{M} 
\end{align}
while the off-diagonal components are given by
\begin{align}
\label{GGMij}
	\frac{(\bG^{T}\bG^{\ast})_{ij}}{M} =
	\sqrt{\beta _{i}\beta _{j}}\frac{Y}{M}.
\end{align}
Expanding $X$ and $Y$ in terms of the complex Gaussian variables, $h_{ij}$, using \eqref{HHMii} and \eqref{HHMij}, it is straightforward to show that: $\mathbb{E}(X)=M$, $\mathbb{E}(Y)=0$. Furthermore, $Var\left\{ \frac{X}{M}\right\} =Var\left\{ \frac{Y}{M}\right\} =\frac{1}{M}$, and thus when $M\rightarrow \infty $ the variance becomes zero. Hence, we have $\mathbb{E}\left\{ \frac{1}{M}(\bG^{T}\bG^{\ast })_{ii}\right\} =\beta _{i}$, $\mathbb{E}\left\{ \frac{1}{M}(\bG^{T}\bG^{\ast })_{ij}\right\} =0$, $\Var\left\{ (\frac{\bG^{T}\bG^{\ast }}{M})_{ii}\right\} =\beta_i/M$ and $\Var\left\{ (\frac{\bG^{T}\bG^{\ast }}{M})_{ij}\right\}=\sqrt{\beta_i\beta_j}/M$. Essentially, the speed of convergence of $\frac{1}{M}\bG^{T}\bG^{\ast }$ is controlled by the convergence of $\bW=\frac{1}{M}\bH^{T}\bH^{\ast }$ to $\bI$ with the $\beta _{j}$ values scaling the variances of the elements of $\bW$.

\vspace{-10pt}
\subsection{Convergence Metrics}
\label{ConvergenceMetrics}
We can further evaluate the convergence of $\bW$ by examining a number of well known properties of $\bW$ and a deviation matrix $\bE=\bW-\bI$. Letting $\lambda _{1},\lambda _{2},\ldots ,\lambda _{K}$ denote the eigenvalues of $\bW$, we consider the following metrics: Mean Absolute Deviation (MAD), $\lambda $ ratio and Diagonal Dominance, defined as
\begin{align}
	\textrm{MAD}(\mathbf{E}) &= \frac{1}{K^{2}}\sum_{i=1,j=1}^{K}|\mathbf{E} _{ij}| , \\
	\lambda \textrm{ ratio} &= \frac{\lambda _{\textrm{max}}(\bW)}{ \lambda_ {\textrm{min}}(\bW)}, \\
	\textrm{Diagonal Dominance} &= \frac{\sum_{i=1}^{K}{\bW_{ii}}}{ \sum_{i=1}^{K}{\sum_{j=1,j\neq i}^{K}{|\bW_{ij}|}}}. \label{diagonaldominance}
\end{align}
These metrics will be evaluated via simulation for a number of system scenarios in Section \ref{sec:Results}.

\subsection{SNR/SINR convergence for ZF and MF precoders}
\label{ZFMFconv}
In this section, we derive limiting expressions for the ZF SNR and MF SINR for the unequal power scenarios. We include a summary of the equal power results \cite{RUSEK} as these are needed for the derivations. As in \cite{RUSEK}, we consider convergence scenario 2.

\subsubsection{Zero Forcing Precoding, Equal Link Gains}
\label{secZFequal}
For a zero forcing precoder, the transmitted symbol vector, $\bsf$ is given by 
\begin{equation}
 \bsf = \frac{1}{\sqrt{\gamma }}\bG^{\ast}(\bG^{T}\bG^{\ast})^{-1}\bqf,
\end{equation}
where we normalize the average power in  $\bsf $ to $\rho _{f}$ via
\begin{equation}
\label{gamZFequal}
\gamma = \frac{\textrm{tr}((\bG^{T}\bG^{\ast})^{-1})}{K}.
\end{equation}
Note that the entries of $\bG$ are iid in this case. The resulting instantaneous received SNR is given by
\begin{equation}
	\textrm{SNR} = \frac{\rho _{f}}{\textrm{tr}((\bG^{T}\bG^{\ast})^{-1})}. \label{snrZFequal}
\end{equation}
The limit of \eqref{snrZFequal} as $M,K\rightarrow \infty $, with fixed $\frac{M}{K}=\alpha $ is \cite{RUSEK}
\begin{equation}
	\textrm{SNR}\rightarrow \rho _{f}(\alpha -1). \label{snrZFequalAsymp} 
\end{equation}

\subsubsection{Zero Forcing Precoding, Unequal Link Gains}
\label{secZFunequal}
Generalizing the analysis in Section \ref{secZFequal} to unequal link gains, we have the power normalization 
\begin{equation}
\label{gamZFunequal}
	\gamma = 
	\frac{\textrm{tr}((\bD _{\beta }^{\frac{1}{2}}\bH ^{T}\bH ^{\ast }\bD _{\beta }^{\frac{1}{2}})^{-1})}{K}.
\end{equation}
Hence, as in \eqref{snrZFequal}, the instantaneous SNR is given 
\begin{align}
\label{SNRzfunequal}
\textrm{SNR} = \frac{\rho _{f}}{\textrm{tr}((\bD _{\beta }^{\frac{1}{2}}\bH ^{T}\bH ^{\ast }\bD _{\beta }^{\frac{1}{2}})^{-1})}.
\end{align}
Considering the denominator of \eqref{SNRzfunequal}, using known properties of the inverse Wishart matrix, we have
\begin{align} 
	\mathbb{E}\left\{ (\bD _{\beta }^{\frac{1}{2}}\bH ^{T}\bH ^{\ast }\bD _{\beta })^{-1}\right\} &= \frac{\bD _{\beta }^{-1}}{M-K}.
\end{align}
Hence,
\begin{align} 
\mathbb{E}\left\{ \textrm{tr}((\bD _{\beta }^{\frac{1}{2}}\bH ^{T}\bH ^{\ast }\bD _{\beta })^{-1})\right\} &= \frac{\sum_{j=1}^{K}{\frac{1}{\beta _{j}}}}{M-K} = \frac{\overline{(\frac{1}{\beta })}}{\alpha -1}, \label{ZFsnrapprox1}
\end{align}
where $\overline{(\frac{1}{\beta })}=\frac{1}{K}\sum_{j=1}^{K}{\beta _{j}^{-1}}$. Assuming that $\overline{(\frac{1}{\beta })}$ converges to the limit, $\overline{(\frac{1}{\beta })^{\infty }}$ as $M,K\rightarrow \infty $ then it can be shown that $\textrm{tr}((\bD_{\beta }^{1/2}\bH^{T}\bH^{\ast }\bD_{\beta }^{1/2})^{-1})\rightarrow \overline{(\frac{1}{\beta })^{\infty }}$. This convergence to the limiting mean follows as the variance vanishes \cite{MUIRHEAD}. Hence, we observe that
\begin{equation}
	\textrm{SNR}\rightarrow \frac{\rho _{f}(\alpha -1)}{\overline{(\frac{1}{\beta })^{\infty }}}. \label{SNRzfunequalAsymp}
\end{equation}

\subsubsection{Matched Filter Precoding, Equal Powers}
\label{secMFequal}
The transmitted signal for a MF precoder is given by 
\begin{equation}
	\bsf = \frac{1}{\sqrt{\gamma }}\bG^{\ast}\mathbf{q}_{f},
\end{equation}
with 
\begin{equation}
\gamma = \frac{\textrm{tr}(\bG^{T}\bG^{\ast})}{K}.
\end{equation}
The received signal is thus given by 
\begin{equation}
  \bxf = \sqrt{\rho _{f}}\bG^{T}\bG^{\ast}\frac{\bqf}{\sqrt{\gamma }} + \bwf,
\end{equation}
giving the instantaneous SINR of the $i$th user as
\begin{equation}
	\textrm{SINR}_i = 
	\frac{\frac{\rho _{f}}{K\gamma }|\bg_{i}^{T}\bg_{i}^{\ast }|^{2}}
	{1 + \frac{\rho _{f}}{K\gamma }\sum_{k=1,k\neq i}^{K}{\bg_{i}^{T}\bg_{k}^{\ast }\bg_{k}^{T}\bg_{i}^{\ast }}}, \label{sinrMFequal}
\end{equation}
where $\bg_{i}$ is the $i$th column of $\bG$. Under the limit operation, as in \cite{RUSEK}, we obtain the limit of \eqref{sinrMFequal} as $M,K\rightarrow \infty $, with $\frac{M}{K}=\alpha $ fixed
\begin{equation}
	\textrm{SINR}_i \rightarrow \frac{\rho _{f}\alpha }{\rho _{f}+1}. \label{sinrMFequalAsymp}
\end{equation}
%


\subsubsection{Matched Filter Precoding, Unequal Powers}
\label{secMFunequal}
Here, the power normalization factor is 
\begin{equation}
\label{gamMFunequal}
	\gamma = 
	\frac{\textrm{tr}(\bD_{\beta }^{\frac{1}{2}}\bH^{T}\bH^{\ast }\bD_{\beta }^{\frac{1}{2}})}{K}.
\end{equation}
%

%
From \eqref{sinrMFequal}, the instantaneous SINR, for unequal powers, of the $i$th user can be shown to be
\begin{align}
	\textrm{SINR}_i &= \frac{\frac{\rho _{f}}{K\gamma }\beta _{i}^{2}|\bh_{i}^{T}\bh_{i}^{\ast }|^{2}}
	{1 + \frac{\rho _{f}\beta _{i}}{K\gamma }\sum_{k=1,k\neq i}^{K}{\beta _{k}|\bh_{i}^{T}\bh_{k}^{\ast }|^{2}}}.
	\label{sinrMFunequal}
\end{align}
In order to examine the asymptotic behaviour of \eqref{sinrMFunequal}, we first rewrite the right-hand side to give
\begin{align}
	\textrm{SINR}_{i} 
			 &= \frac{\frac{\rho _{f}(\frac{M}{K})\beta _{i}^{2}}{\frac{\gamma }{M}}|\frac{\bh_{i}^{T}\bh_{i}^{\ast }}{M}|^{2}}{1 + \frac{\rho _{f}\beta _{i}}{\frac{\gamma }{M}}\left( \frac{\sum_{k=1,k\neq i}^{K}{\beta _{k}}}{K}\right) \left( \frac{\sum_{k=1,k\neq i}^{K}{|\bh_{i}^{T}\bh_{k}^{\ast }|^{2}\beta _{k}}}{M\sum_{k=1,k\neq i}^{K}{\beta _{k}}}\right) }.
			\label{sinrMFunequaExpanded}
\end{align}
We note the following properties of the terms in \eqref{sinrMFunequaExpanded}. In the numerator, $\frac{\bh_{i}^{T}\bh_{i}^{\ast }}{M}\rightarrow 1$ as $M\rightarrow \infty$. 
Also, we have 
\begin{align}
	\frac{\gamma }{M} &= \frac{\textrm{tr}(\bD_{\beta }^{\frac{1}{2}}\bH^{T}\bH^{\ast }\bD_{\beta }^{\frac{1}{2}})}{MK} \notag \\
	&\rightarrow \lim_{K\rightarrow \infty }\frac{\textrm{tr}(\bD_{\beta })}{K} \notag \\
	&= \lim_{K\rightarrow \infty }\frac{\sum_{k=1}^{K}{\beta _{k}}}{K} = \overline{\beta ^{\infty }}, \label{gamM}
\end{align}
assuming that the limit $\frac{1}{K}\sum_{k=1}^{K}{\beta _{k}}\rightarrow \overline{\beta ^{\infty }}$ exists. Note that if the limit, $\overline{\beta ^{\infty }}$, exists then in \eqref{sinrMFunequaExpanded}, the terms $\frac{1}{K}\sum_{k=1,k\neq i}^{K}{\beta _{k}}$ will also converge so that
\begin{equation}
	\frac{1}{K}\sum_{k=1,k\neq i}^{K}{\beta _{k}}\rightarrow \overline{\beta ^{\infty }}.
\end{equation}
%

%
%
%
%
%
%

%
Finally, from Sec. \ref{GGconv} and \eqref{gamM}, we have
\begin{equation}
	\frac{1}{MK}\sum_{k=1,k\neq i}^{K}{|\bh_{i}^{T}\bh_{k}^{\ast }|^{2}\beta _{k}}\rightarrow \overline{\beta ^{\infty}}. \label{star}
\end{equation}
Combining \eqref{gamM} and \eqref{star} with \eqref{sinrMFunequaExpanded} we have the limiting SINR of \eqref{sinrMFunequal} for $M,K\rightarrow \infty $, with fixed $\frac{M}{K}=\alpha $, for user $i$ given by
\begin{align}
	\textrm{SINR}_i
			 &\rightarrow \frac{\frac{\rho _{f}\alpha \beta _{i}^{2}\times 1}{\overline{\beta ^{\infty }}}}{1 + \frac{\rho _{f}\beta _{i}}{\overline{\beta ^{\infty }}}\overline{\beta ^{\infty }}\times 1} 
			 = \frac{\rho _{f}\alpha \beta _{i}^{2}}{\overline{\beta ^{\infty }}+\rho _{f}\beta _{i}\overline{\beta ^{\infty }}}. \label{SINRmfunequalAsymp}
\end{align}

\section{Results}
\label{sec:Results}
We consider the two convergence scenarios given in Sec. \ref{ConvergenceModel} and the convergence metrics given in Sec. \ref{ZFMFconv}
\subsection{Convergence Properties}
%
\begin{figure}[ht]
\centering\includegraphics[width=1\columnwidth]{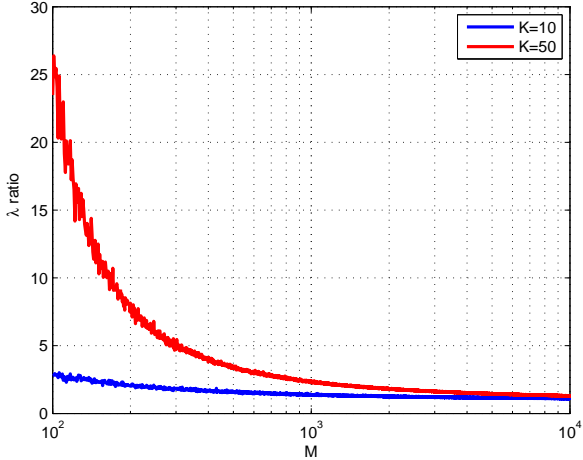}
\vspace{-5pt}
\raisecaption\caption{$\lambda $ ratio vs $M$ for an iid channel with $K$ fixed}
\label{fig1}
\end{figure}
%
\begin{figure}[ht]
\vspace{-15pt}
\centering\includegraphics[width=1\columnwidth]{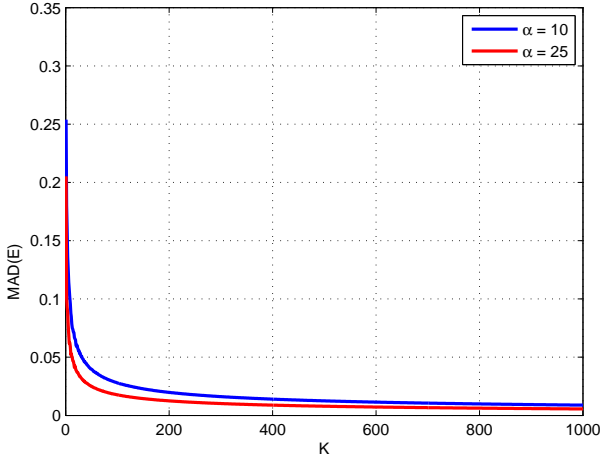}
\vspace{-5pt}
\raisecaption\caption{$\textrm{MAD}(\bE)$ vs $K$ for an iid channel with $\alpha =M/K$ fixed}
\label{fig2}
\end{figure}
%
\begin{figure}[ht]
\centering\includegraphics[width=1\columnwidth]{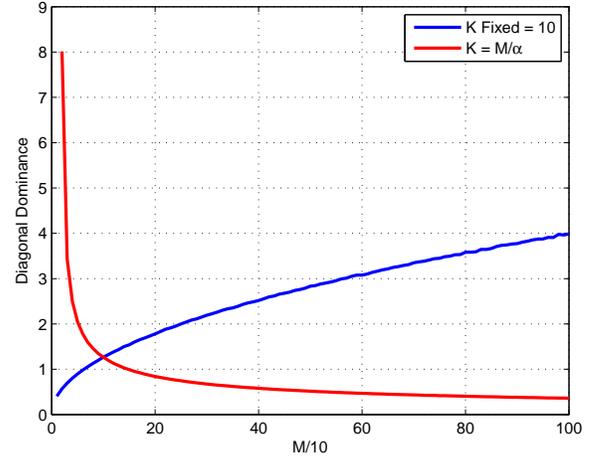}
\raisecaption\caption{iid Channel Diagonal Dominance, both $K$ fixed and $\frac{K}{M}=\alpha^{-1}$ for $M$ increasing}
\label{fig3}
\end{figure}
Fig. \ref{fig1} shows the $\lambda $ ratio versus $M$ for convergence scenario 1 with $K=10$ and $K=50$. It can be seen that for both values of $K$, the $\lambda $ ratio will only converge to 1 when $M$ is excess of $10^{4}$. Even for values of $M$ as large as 100, the $\lambda $ ratio for $K=50$ is more than 8 time larger than the $\lambda $ ratio for $K=10$. However, when we consider Fig. \ref{fig2} showing the mean absolute deviation of $\bH^{T}\bH^{\ast }/M$ from the identity $\bI$, we note that this difference quickly approaches zero. For example, in Fig. \ref{fig1}, for $M = 500$ the $\lambda $ ratio is about 4 for $K=50$, yet in Fig. \ref{fig2}, when $K=50$, and the corresponding $M=500$, the mean absolute deviation is less than 0.05.
\par

In Fig. 3, we observe that $\bW$ becomes increasingly diagonal dominant for fixed $K$ as $M\rightarrow \infty $. This is because $\bW$ has fixed dimension ($K\times K$) and the sum of the diagonal elements grow faster thank the fixed number of off-diagonals. In contrast as both $M$ and $K$ grow large, $\bW$ becomes less diagonally dominant. This follows as the number of off-diagonal elements increases as $K^{2}$ and the total contribution of the off-diagonals becomes dominant. It can be shown that the diagonal dominance measure grows proportionally to $M^{1/2}$ for fixed $K$ and decays proportionally to $M^{-1/2}$ as both $M$ and $K$ increase. For reasons of space, details are omitted.

\par
Considering Figs. \ref{fig1} and \ref{fig2}, we may conclude that the massive MIMO behaviour that results in a deterministic $\bH^{T}\bH^{\ast }/M$ only begins to show at some very large values of $M$ - the base station antenna numbers.

\subsection{Convergence Properties of ZF and MF precoders}

\begin{figure}[ht]
\centering\includegraphics[width=1\columnwidth]{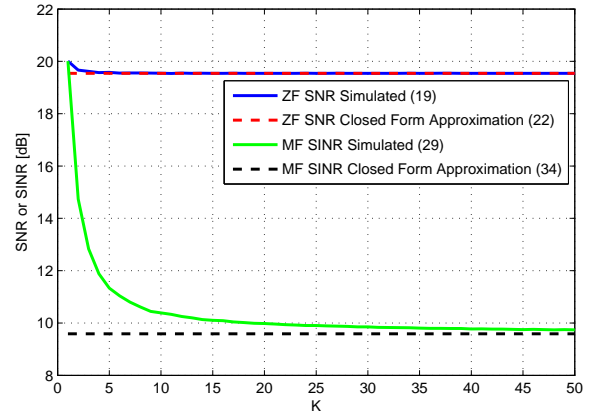}
\raisecaption\caption{ZF SNR and MF SINR, Single Antenna MU-MIMO, Equal Powers, $\frac{K}{M}=\alpha^{-1}=\frac{1}{10}$}
\label{fig4}
\end{figure}

\begin{figure}[ht]
\centering\includegraphics[width=1\columnwidth]{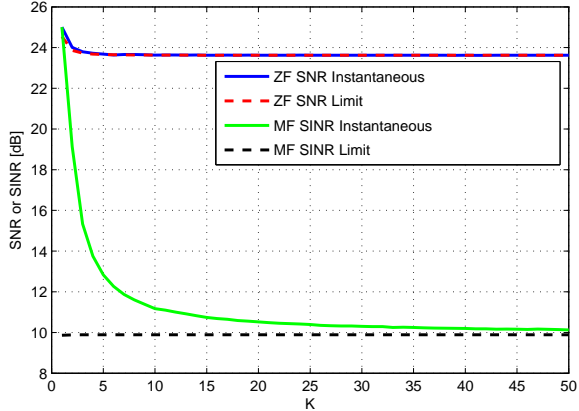}
\raisecaption\caption{ZF SNR and MF SINR, Single Antenna MU-MIMO, Unequal Powers, $\frac{K}{M}=\alpha^{-1}=\frac{1}{10}$}
\label{fig5}
\end{figure}

Figs. \ref{fig4} and \ref{fig5} show the convergence properties of the ZF and MF precoders for equal and unequal powers, respectively, where we plot the expected value of \eqref{snrZFequal} and \eqref{SNRzfunequal} for ZF and \eqref{sinrMFequal} and \eqref{sinrMFunequal} for MF, along with their corresponding limits. The ZF precoder in both cases of equal and unequal powers converges to the limit quickly. For example, in Figs. \ref{fig4} and \ref{fig5}, when $K=10$ and $M=100$ the expected value of the per-user SNR already approaches the asymptotic limit for infinite antennas. 
The convergence of the MF precoding is shown also in Figs. \ref{fig4} and \ref{fig5}. It can be observed in both figures that the SINR for the MF precoder is not equal to the SNR of the ZF precoder in the asymptotic limits; the MF precoder SINR is effectively reduced by a factor of $\rho _{f}+1$ when $\alpha $ is large (looking at \eqref{snrZFequalAsymp} and \eqref{sinrMFequalAsymp}), as compared to the ZF SNR. This difference is obvious in Figs. \ref{fig4} and \ref{fig5} even for $\alpha =10$. The MF SINR takes a longer time to converge because of the additional random variables in the numerator and denominator of \eqref{sinrMFequalAsymp}. The unequal power case for the MF precoder has additional terms in \eqref{SINRmfunequalAsymp} which manifest itself in a small increase of the per-user SINR as compared to the equal powers. The MF precoder SINR is also less relative to ZF SNR due to the inter-user interference terms in the denominator of \eqref{sinrMFequal}; the boost in the SINR in the numerator of \eqref{sinrMFequal} due to the co-phasing terms $|\bg_{i}^{T}\bg_{i}^{\ast }|^{2}$ is not enough to compensate for the inter-user interference given by $\sum_{k=1,k\neq i}^{K}{\bg_{i}^{T}\bg_{k}^{\ast }\bg_{k}^{T}\bg_{i}^{\ast }}$.

\subsection{Impact of Correlation}

\begin{figure}[ht]
\centering\includegraphics[width=1\columnwidth]{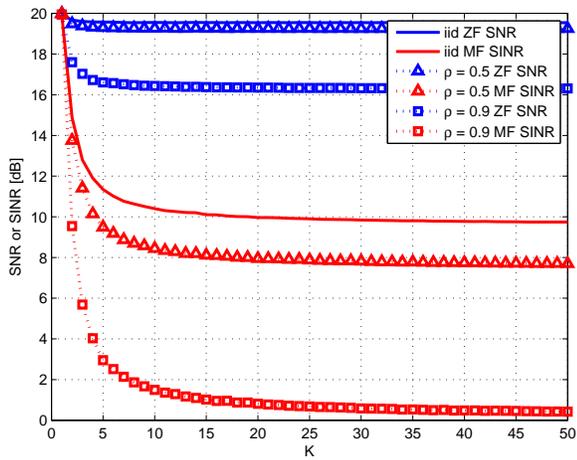}
\raisecaption\caption{ZF SNR and MF SINR, Single Antenna MU-MIMO, Correlated Channel, Equal Powers, $\frac{K}{M}=\alpha^{-1}=\frac{1}{10}$}
\label{fig6}
\end{figure}

\begin{figure}[ht]
\centering\includegraphics[width=1\columnwidth]{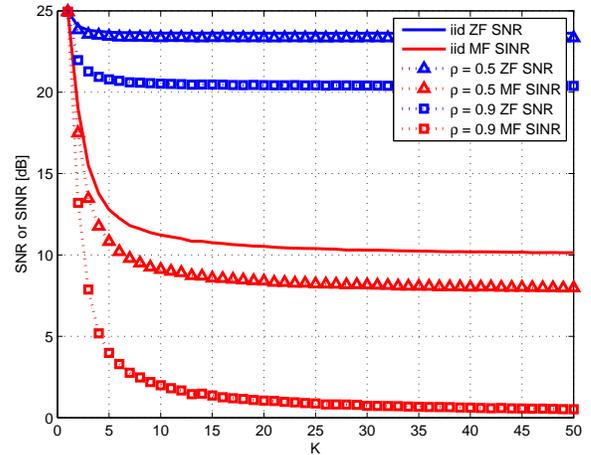}
\raisecaption\caption{ZF SNR and MF SINR, Single Antenna MU-MIMO, Correlated Channel, Unequal Powers, $\frac{K}{M}=\alpha^{-1}=\frac{1}{10}$}
\label{fig7}
\end{figure}

Correlation is introduced in \eqref{GHDdef} by using the Kronecker model \eqref{Kronecker}. In the simulations two cases of correlation are shown, a high inter-element correlation of $\rho =0.9$ and a low inter-element correlation of $\rho =0.5$. Figs. \ref{fig6} and \ref{fig7} show the effect of this correlation for ZF and MF precoders for the mean SNR and SINR respectively. We can see that correlation includes a large penalty in the mean per-user SNR and SINR for both ZF and MF precoders (equal and unequal powers), when compared to the corresponding uncorrelated case. The SNR/SINRs for the lower value of correlation should be similar to the i.i.d. case for the two precoders - and this is indeed the case. 

\section{Conclusion}
\label{sec:Conclusion}
In this paper, we have analyzed the number of antennas required for massive MIMO properties. We have presented a method to derive ZF SNR and MF SINR for a scenario of unequal powers amongst the users. We have also derived the limit of these SNRs/SINRs as the number of base station antennas is increased indefinitely. We found that massive MIMO property that relies on small values of the off-diagonal elements of $\bH^{T}\bH^{\ast }/M$ is desired, then a relatively small number of base station antennas will achieve convergence. On the other hand if the desirable matrix is to reduce the spread between maximum and minimum eigenvalues, then much larger base station antennas are needed. 
\par
Interestingly, the per-user SNR/SINR for both ZF and MF precoers are less sensitive to $M$. In particular the per-user SNR for a ZF precoder converges quickly even for small values of $M$ even though the SNR expression requires the computation of the inverse of a matrix. This leads us to conclude that for ZF and MF precoders, we can use the limiting value of SNR/SINR even for small values of $M$. Also, a correlated channel reduces MF SINR by a significant amount as compared with ZF SNR.


\bibliographystyle{IEEEtran}
\bibliography{bibliography}

\begin{thebibliography}{10}
\providecommand{\url}[1]{#1}
\csname url@samestyle\endcsname
\providecommand{\newblock}{\relax}
\providecommand{\bibinfo}[2]{#2}
\providecommand{\BIBentrySTDinterwordspacing}{\spaceskip=0pt\relax}
\providecommand{\BIBentryALTinterwordstretchfactor}{4}
\providecommand{\BIBentryALTinterwordspacing}{\spaceskip=\fontdimen2\font plus
\BIBentryALTinterwordstretchfactor\fontdimen3\font minus
  \fontdimen4\font\relax}
\providecommand{\BIBforeignlanguage}[2]{{%
\expandafter\ifx\csname l@#1\endcsname\relax
\typeout{** WARNING: IEEEtran.bst: No hyphenation pattern has been}%
\typeout{** loaded for the language `#1'. Using the pattern for}%
\typeout{** the default language instead.}%
\else
\language=\csname l@#1\endcsname
\fi
#2}}
\providecommand{\BIBdecl}{\relax}
\BIBdecl

\bibitem{LARSSON}
E.~G. Larsson, F.~Tufvesson, O.~Edfors, and T.~L. Marzetta, ``Massive {MIMO}
  for next generation wireless systems,'' \emph{IEEE Communications Magazine},
  April 2013.

\bibitem{MARZETTA}
T.~L. Marzetta, ``Noncooperative cellular wireless with unlimited numbers of
  base station antennas,'' \emph{IEEE Transactions on Wireless Communications},
  vol.~9, no.~11, pp. 3590--3600, November 2010.

\bibitem{NGO}
H.~Q. Ngo, E.~G. Larsson, and T.~L. Marzetta, ``Energy and spectral efficiency
  of very large multiuser {MIMO} systems,'' \emph{IEEE Transactions on
  Communications}, vol.~61, no.~4, pp. 1436--1149, April 2013.

\bibitem{PITAROKOILIS}
A.~Pitarokoilis, S.~K. Mohammed, and E.~G. Larsson, ``On the optimality of
  single-carrier transmission in large-scale antenna systems,'' \emph{IEEE
  Wireless Communications Letters}, vol.~1, no.~4, pp. 276--279, August 2012.

\bibitem{HOYDIS}
J.~Hoydis, S.~ten Brink, and M.~Debbah, ``Massive {MIMO} in the {UL}/{DL} of
  cellular networks: How many antennas do we need?'' \emph{IEEE Journal on
  Selected Areas in Communications}, vol.~31, no.~2, pp. 160--171, February
  2013.

\bibitem{MULLER}
R.~R. Muller, M.~Vehkapera, and L.~Cottatellucci, ``Blind pilot
  decontamination,'' \emph{International ITG Workshop on Smart Antennas}, March
  2013.

\bibitem{MOUSTAKAS}
A.~Moustakas, S.~Simon, and A.~Sengupta, ``{MIMO} capacity through correlated
  channels in the presence of correlated interferers and noise: a (not so)
  large n analysis,'' \emph{IEEE Transactions on Information Theory}, vol.~49,
  no.~10, pp. 2545--2561, October 2003.

\bibitem{CHUAH}
C.-N. Chuah, D.~N.~C. Tse, J.~M. Kahn, and R.~A. Valenzuela, ``Capacity scaling
  in {MIMO} wireless systems under correlated fading,'' \emph{IEEE Transactions
  on Information Theory}, vol.~48, no.~2, pp. 637--650, February 2002.

\bibitem{RUSEK}
F.~Rusek, D.~Persson, B.~K. Lau, E.~G. Larsson, T.~L. Marzetta, O.~Edfors, and
  F.~Tufvesson, ``Scaling up {MIMO}: Opportunities and challenges with very
  large arrays,'' \emph{IEEE Signal Processing Magazine}, vol.~30, no.~1, pp.
  40--60, January 2013.

\bibitem{LOYKA}
S.~L. Loyka, ``Channel capacity of {MIMO} architecture using the exponential
  correlation matrix,'' \emph{IEEE Communications Letters}, vol.~5, no.~9, pp.
  369--371, September 2001.

\bibitem{PAPOULIS}
A.~Papoulis and S.~U. Pillai, \emph{Probability, Random Variables And
  Stochastic Processes}, 4th~ed.\hskip 1em plus 0.5em minus 0.4em\relax
  McGraw-Hill Europe, 2002.

\bibitem{MUIRHEAD}
R.~J. Muirhead, \emph{Aspects of Multivariate Statistical Theory},
  2nd~ed.\hskip 1em plus 0.5em minus 0.4em\relax Wiley, 2005.

\end{thebibliography}

\end{document}